# NEW $BVI_C$ PHOTOMETRY OF LOW-MASS PLEIADES STARS: EXPLORING THE EFFECTS OF ROTATION ON BROADBAND COLORS

Brittany L. Kamai[1], Frederick J. Vrba[2], John R. Stauffer[3], and Keivan G. Stassun[1,4]
[1] Department of Physics & Astronomy, Vanderbilt University, VU Station B 1807, Nashville, TN 37235, USA
[2] U.S. Naval Observatory, 10391 West Naval Observatory Road, Flagstaff, AZ 86001-8521, USA
[3] Spitzer Science Center, California Institute of Technology, MC 220-6, Pasadena, CA 91125, USA
[4] Fisk University, 1000 17th Avenue North, Nashville, TN 37208, USA

## ABSTRACT

We present new $BVI_C$ photometry for 350 Pleiades proper motion members with $9 < V \lesssim 17$. Importantly, our new catalog includes a large number of K- and early M-type stars, roughly doubling the number of low-mass stars with well-calibrated Johnson/Cousins photometry in this benchmark cluster. We combine our new photometry with existing photometry from the literature to define a purely empirical isochrone at Pleiades age ($\approx$100 Myr) extending from $V = 9$ to 17. We use the empirical isochrone to identify 48 new probable binaries and 14 likely nonmembers. The photometrically identified single stars are compared against their expected positions in the color–magnitude diagram (CMD). At 100 Myr, the mid K and early M stars are predicted to lie above the zero-age main sequence (ZAMS) having not yet reached the ZAMS. We find in the $B - V$ versus $V$ CMD that mid K and early M dwarfs are instead displaced below (or blueward of) the ZAMS. Using the stars' previously reported rotation periods, we find a highly statistically significant correlation between rotation period and CMD displacement, in the sense that the more rapidly rotating stars have the largest displacements in the $B - V$ CMD.

*Key words:* catalogs – Hertzsprung–Russell and C–M diagrams – open clusters and associations: individual (Pleiades) – stars: low-mass – stars: rotation – starspots

*Online-only material:* color figures, machine-readable and VO table

## 1. INTRODUCTION

The Pleiades cluster is one of the quintessential open clusters that defines the empirical isochrone of solar metallicity stars at ~100 Myr. Empirical isochrones are derived from the single star locus of different age clusters. They serve to both utilize evolutionary model predictions and to point the way to needed improvements in those models. The Pleiades serves as an ideal testing ground by virtue of its proximity and richness (133 pc, Soderblom et al. 2005), membership (1400 members, Stauffer et al. 2007), and solar metallicity (Soderblom et al. 2009). It is old enough and rich enough to allow determination of a reasonably accurate upper-main-sequence (UMS) turn-off age. At the same time, it is young enough, in principle, to allow the determination of a reasonably accurate pre-main-sequence (PMS) age. Current models do well at matching most broadband observations at wavelengths longer than 1 micron. However, accurate atmosphere models are a limiting factor to transforming theoretical predictions into direct observables for shorter wavelengths. Empirical isochrones from observed, fiducial open clusters are employed to fine tune model predictions. Any inconsistencies between observations and predictions will motivate further modeling to understand the underlying physical mechanisms.

Standard models (e.g., Baraffe et al. 1998) predict that at 100 Myr, K and M dwarfs will have not yet contracted onto the ZAMS and would still be in their PMS phase. A signature of this should be a systematic shift of the K and M stars above the ZAMS in an H–R diagram or color–magnitude diagram (CMD). The theoretical ZAMS (Baraffe et al. 1998), theoretical 100 Myr (Baraffe et al. 1998), semi-empirical ZAMS (VandenBerg & Clem 2003), and the previously derived empirical 100 Myr sequence (Stauffer et al. 2007) are compared in Figure 1. This highlights the differences between the theoretical predictions and empirical observations.

A striking feature is that the empirical 100 Myr isochrone (derived from the Pleiades) exhibits a systematic shift below (or blueward of) the nominal ZAMS in the $B - V$ versus $V$ CMD only. Observations of this blueward departure in the Pleiades dates back almost 50 yr. Explanations of this peculiar behavior range from a non-coeval formation history (Herbig 1962) to obscuration by dust cocoons (Jones 1972).

Most recently, Stauffer et al. (2003) further investigated the low-mass Pleiads anomalous blueward displacement for the K dwarfs in the $B - V$ versus $V$ CMD. The authors also found a redward displacement in the redder colors such as in the $V - K$ versus $V$ CMD. Stauffer et al. (2003) proposed that the observed trend in the CMDs is from extensive spot and plage coverage on the low-mass stars. Hot plages can make the $B - V$ colors too blue, cool spots might simultaneously make the $V - K$ colors too red, and the overall effect minimized in the intermediate $V - I_C$ color. Those authors further speculated that if this spot/plage activity effect is correct, then in fact, stellar rotation may be the cause.

Studies of the rotational properties of Pleiades stars have a long history. With the advent of CCD-era photometry, numerous studies sought to make use of spot-modulated brightness variations to directly determine the stellar rotation periods (for a summary of the state of the field prior to 2000, see the volume of the ASP Conference Series 198, Pallavicini et al. 2000, in particular the contributions of Stauffer et al., Stassun et al., and Barnes et al.). Until recently, however, there were relatively few stars in the Pleiades, especially at low masses, with successfully measured rotation periods from spot modulations. Fortunately, rotation periods for a large number of Pleiades members has recently become possible with wide-field photometric

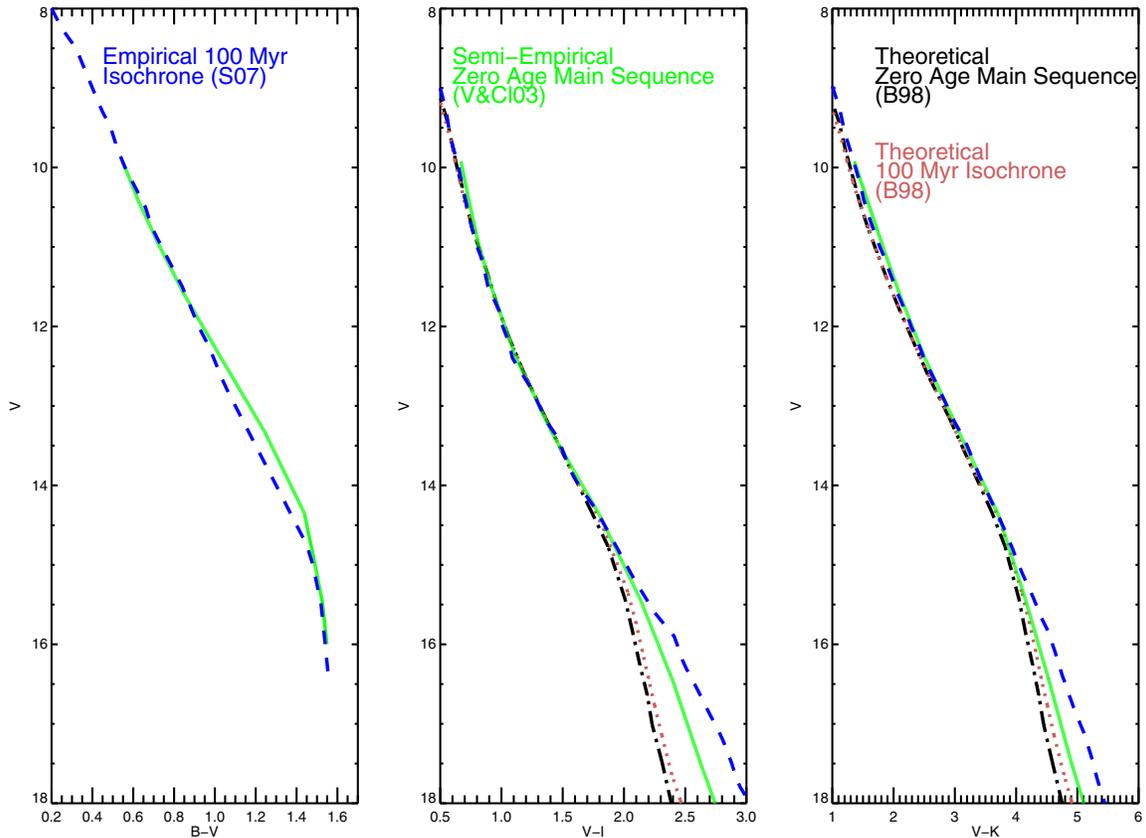

**Figure 1.** Comparison of theoretical ZAMS (black dash–dotted line), semi-empirical ZAMS (green solid line), theoretical 100 Myr (brown dotted line), and empirical 100 Myr (blue dashed line) isochrones. The theoretical isochrones from Baraffe et al. (1998) use solar metallicity as appropriate for the Pleiades (Fe/H = 0.03 ± 0.02, Soderblom et al. 2009). Semi-empirical isochrones are from VandenBerg & Clem (2003, V03), who converted the theoretical isochrones into the observational plane using empirical color–$T_{\rm eff}$ relations matched to open and globular cluster photometry. The empirical isochrones are from Stauffer et al. (2007) based on available Pleiades photometry.

(A color version of this figure is available in the online journal.)

monitoring surveys. In particular, Hartman et al. (2010) have recently reported photometric rotation periods for 368 Pleiads, and in this paper we make use of these new rotation periods to directly explore the connection between rotation and the stellar color displacements discussed above.

A main motivation of this work is to continue an investigation of the physical processes that could cause anomalous color behavior of the low-mass Pleiades (mid K to early M), particularly to test the hypothesis of rotation-driven spot/plage effects. In Section 2, we report new homogeneous photometry in the $BVI_C$ passbands for 350 Pleiades members, approximately doubling the number of low-mass stars with well-calibrated optical photometry. In Section 3.1, we combine this photometry with existing photometry from the literature to determine a new set of empirical isochrones for the Pleiades based on a richer sample, especially within a range of $9 < V \lesssim 17$ (encompassing F5 - M3 dwarfs). In Section 3.2, we use this newly defined empirical single-star isochrone to identify likely binaries, and we moreover extend the work of Stauffer et al. (2003) to further characterize the nature of the color displacements between low-mass Pleiades single stars and a semi-empirical ZAMS. Connecting our color displacement measurements with rotation period measurements from the literature, Section 3.4 presents the result of our test of the rotation hypothesis, where we verify a highly statistically significant correlation between rotation period and displacement such that the most rapidly rotating stars are the most displaced in the CMD. Section 5 concludes with a summary of our findings.

## 2. DATA

### 2.1. New Observations

We targeted 350 proper motion Pleiades members within a range of $9 < V \lesssim 17$ for new homogeneous $BVI_c$ observations (Table 1). Target stars were selected from the Stauffer et al. (2007, S07) catalog, which compiles all the photometry from the literature in the $BVI_C JHK$ photometric bands for the ∼1400 proper motion members. The 350 targets were selected by virtue of being the brightest $K$-band proper motion members that had either no, incomplete, or outdated (photographic plate) $BVI_C$ photometry.

U.S. Naval Observatory optical observations were obtained over 36 nights from 2007 November 11 to 2009 January 31 with the 1.0 m Ritchey–Chretien telescope at the Flagstaff Station using a Tektronix 1024x1024 thinned, backside-illuminated CCD. The standard Johnson $B$ and $V$ filters and a Kron–Cousins $I_C$ filter were employed. At the beginning of each night, bias frames and dome-flat frames for each filter were obtained. Each target frame was de-biased and flat-field corrected in real time. Typically, 12–15 photometric standard field stars were observed throughout each night, with anywhere from 1 to 10 standards in each field. The standards were selected from the lists of Landolt (1983a, 1983b, 1992) to cover the expected color range of the stars in this study. All standard stars in each field were used to determine nightly atmospheric extinction coefficients in each filter. The accuracies of the resulting extinction coefficients

**Table 1**
New $BVI_C$ Pleiades Photometry

| R.A. (J2000.0) | Decl. (J2000.0) | V | $\sigma V$ | $B-V$ | $\sigma B-V$ | $V-I$ | $\sigma V-I$ | No. Obs | Flag | Name |
|---|---|---|---|---|---|---|---|---|---|---|
| 3 56 01.44 | 23 07 35.04 | 8.763 | 0.023 | 0.360 | 0.019 | 0.360 | 0.019 | 1 | | DH845 |
| 3 56 52.73 | 20 05 00.96 | 9.123 | 0.023 | 0.540 | 0.019 | 0.645 | 0.019 | 1 | b | TrS183x;HD24665 |
| 3 37 24.05 | 22 21 03.60 | 9.223 | 0.023 | 0.502 | 0.019 | 0.609 | 0.019 | 1 | a | PELS003;HD22444;TrS19 |
| 3 45 01.68 | 19 33 33.84 | 9.380 | 0.019 | 0.451 | 0.016 | 0.578 | 0.018 | 2 | | PELS086 |
| 4 03 44.16 | 22 56 39.48 | 9.432 | 0.023 | 0.486 | 0.019 | 0.580 | 0.019 | 1 | | AKIII79 |
| 3 58 01.70 | 20 40 36.48 | 9.449 | 0.023 | 0.463 | 0.019 | 0.617 | 0.019 | 1 | | PELS135 |
| 3 43 50.66 | 25 16 08.04 | 9.449 | 0.023 | 0.497 | 0.019 | 0.569 | 0.019 | 1 | | PELS140;HD23312;TrS61 |
| 3 49 52.92 | 25 38 51.00 | 9.556 | 0.023 | 0.487 | 0.019 | 0.611 | 0.019 | 1 | | PELS006;TrS4 |
| 3 29 25.39 | 25 39 08.28 | 9.573 | 0.023 | 0.500 | 0.019 | 0.552 | 0.019 | 1 | | PELS025 |
| 3 34 07.32 | 24 20 40.20 | 9.579 | 0.023 | 0.507 | 0.019 | 0.634 | 0.019 | 1 | | PELS150;HD23935;TrS144y |

**Notes.**
[a] Binary identified in all CMDs.
[b] Binary identified in 3 out of 4 CMDs.
[c] Probable nonmember.
**References.** Stellar identification nomenclature is taken from Stauffer et al. (2007). A: (Asiago flare star) see Haro et al. (1982). AK: Artyukhina & Kalinina (1970). BPL: (Burrell Pleiades) Pinfield et al. (2000). DH: Deacon & Hambly (2004). H II: Hertzprung (1947). HCG: Haro et al. (1982). HHJ: Hambly et al. (1991). MT: McCarthy & Treanor (1964). PELS: (Pels proper motion member) see van Leeuwen et al. (1986). SK: Stauffer, Klemola et al. (1991). Tr: Trumpler (1921).

(This table is available in its entirety in machine-readable and Virtual Observatory (VO) forms in the online journal. A portion is shown here for guidance regarding its form and content.)

were typically 0.01 mag airmass$^{-1}$ with a few poorer nights having ∼0.015 mag airmass$^{-1}$. In no case were the determined extinction coefficients significantly different than the nominal Flagstaff Station coefficients determined over the past 40 years and in all the cases the nightly coefficients were adopted.

Aperture photometry was carried out using the DAOPHOT "phot" package in IRAF. A single aperture size appropriate for the worst-seeing frames was selected after inspection of all frames for each night. Typically, the worst-seeing frames were those taken at high air mass for the extinction measurements. This conservative approach has the advantage of minimizing systematic magnitude offsets at the cost of slightly increasing the aperture photometric errors.

After removal of atmospheric extinction, $B-V$ and $V-I$ color transformations were determined for each night's data. The CCD and filters bandpasses were well-matched to the standard Johnson and Kron–Cousins systems. Confirmation of consistency requires that the derived transformation slopes be close to 1.0 and the measurements of $B-V$ and $V-I$ transformation slopes were 1.015 and 0.990, respectively. The resulting transformations were then applied to the program star instrumental magnitudes to produce the final photometric results.

The quoted uncertainties are based on the CCD read noise and photon statistics convolved with the color and magnitude transformation errors each night. The Tektronix 1024 x 1024 CCD has a gain of 7.4 $e^-$/ADU, read noise of 8 $e^-$, and bias stability of better than 1 ADU. Typical $V$, $B-V$ and $V-I$ rms values are 0.023, 0.019, and 0.019, respectively. Extinction coefficients were not explicitly included as they are considered to be convolved into the transformation errors.

Sixty-one stars were observed on multiple nights (59 were observed on two nights and 2 were observed on three nights). The photometry reported in Table 1 for these multiple observations is the average of the observations. We used these multiply observed stars to provide an upper limit to the true internal uncertainties of the new photometry. Errors are estimated by $\sigma_{\rm int} = ({\rm MAD}/1.10)$, where MAD is the mean absolute deviation. The rms values of $V$, $B-V$ and $V-I$ are 0.036, 0.024, and 0.21 mag, respectively. These estimates reflect an upper limit on the true photometric errors because of the intrinsic variability of young low-mass stars.

### 2.2. Data from the Literature

To supplement the new observations, 507 cluster members with $BVI_C$ photometry have been selected from the S07 catalog. To produce a homogeneous data set, measurements obtained with photographic plates and filter sets lacking proper calibration to the Johnson–Cousins and Kron systems are excluded. Figure 2 compares the newly obtained and previously published works. The new and previous measurements are largely consistent with one another, forming a tight single-star main sequence between $8 < V \lesssim 17$, including a clear binary sequence elevated above the single-star main sequence.

## 3. RESULTS

In this section, we present the results of our new $BVI_C$ photometric catalog for the Pleiades proper motion members, combining our newly obtained photometry with high-quality literature photometry. First, we use the new photometric catalog to refine the empirical 100 Myr isochrone in the $B-V$, $V-I_C$, and $V-K$ CMDs. Second, we identify likely multiple-star systems in the catalog by virtue of their displacement above the newly defined empirical single-star sequence. Third, we compare the newly defined empirical single-star sequences to the semi-empirical ZAMS in order to reveal the nature of the Pleiades color displacements in the three CMDs. In Section 3.4, we discuss the relation of these color displacements to stellar rotation.

### 3.1. Refining the Empirical 100 Myr Isochrone

Following the procedure in S07, isochrones in each CMD are defined using the single star cluster population. Given a large enough population of stars, the single-star locus can be defined as the mode of the distribution in a CMD, assuming a significant fraction of the stars are single or have binary companions with sufficiently small flux ratios. Even with 350 new stars, there are too few stars in a given color bin to allow for an automated mode-following curve to be defined, therefore the pattern recognition capability of the human eye is used. An iterative process described below is employed to refine the 100 Myr isochrone and identify binaries and nonmembers from the new photometric catalog. The initial guess preferentially fits the lower envelope of the locus since binaries are typically brighter than the single-star locus.

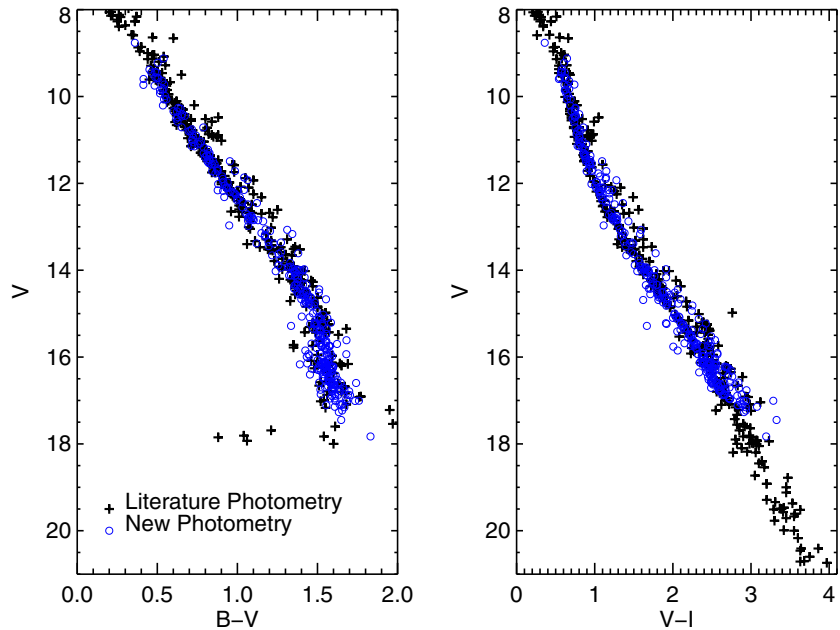

**Figure 2.** 350 stars with newly obtained $BVI_C$ photometry (blue circles) together with 507 stars with reliable measurements from the literature (black cross) form the basis for our refinement of the empirical 100 Myr isochrones in three different color–magnitude planes.
(A color version of this figure is available in the online journal.)

Figure 3 represents the "by eye" fitting method of simultaneously comparing a color–magnitude diagram with a color–displacement diagram to ensure a symmetric scatter (for the lower envelope of stars) about the newly defined isochrone. Each star's $V$ magnitude displacement relative to the isochrone is defined as $\Delta V = V_{obs} - V_{ms}$, where $V_{obs}$ is the observed $V$ magnitude and $V_{ms}$ is the newly defined empirical sequence $V$ magnitude at a given observed color. Measurement errors and stellar variability will scatter stars above and below the isochrone resulting in either positive or negative $\Delta V$ values. A conservative $3\sigma$ offset error of $\Delta V \sim 0.3$ mag is based on a typical photometric error of 0.024 mag in $B - V$ color and the slope of the $B - V$ curve, which is $\sim$5. For the entire study, $\Delta V$ values calculated in the $B - V$ CMD for stars with $B - V$ color values greater than 1.45 are excluded. This is because the main sequence in the $B - V$, CMD becomes nearly vertical (i.e., independent of $B - V$), fainter than $V \sim 15.5$, and small errors in the $B - V$ photometry would result in large apparent displacements relative to an assumed ZAMS.

Similar procedures were employed for the $V - I$ and $V - K$ CMDs. The new empirical isochrones reported in Table 2 are now based on a robust set of high-quality $BVI_C$ photometry for all known, probable Pleiades members with $V \lesssim 17$. The difference between the new isochrones and those of S07 differ by $\sim$0.1 mag at most in $V$. This improved isochrone enables an exploration of the trend of CMD displacement versus rotation period which, as we show in Section 3.4, is at the level of $\sim$0.1–0.3 mag.

**Table 2**
100 Myr Isochrone

| $V$ | $B - V$ | $V - I_c$ | $V - K$ |
|---|---|---|---|
| 9.000 | 0.400 | 0.500 | 1.000 |
| 9.250 | 0.430 | 0.540 | 1.100 |
| 9.500 | 0.480 | 0.570 | 1.220 |
| 9.750 | 0.520 | 0.610 | 1.300 |
| 10.000 | 0.550 | 0.650 | 1.380 |
| 10.250 | 0.600 | 0.690 | 1.470 |
| 10.500 | 0.650 | 0.730 | 1.550 |
| 10.750 | 0.690 | 0.770 | 1.660 |
| 11.000 | 0.740 | 0.810 | 1.780 |
| 11.250 | 0.790 | 0.840 | 1.900 |
| 11.500 | 0.830 | 0.890 | 2.000 |
| 11.750 | 0.890 | 0.950 | 2.140 |
| 12.000 | 0.920 | 1.010 | 2.280 |
| 12.250 | 0.970 | 1.070 | 2.390 |
| 12.500 | 1.030 | 1.130 | 2.560 |
| 12.750 | 1.070 | 1.220 | 2.720 |
| 13.000 | 1.120 | 1.300 | 2.870 |
| 13.250 | 1.180 | 1.370 | 3.030 |
| 13.500 | 1.240 | 1.440 | 3.160 |
| 13.750 | 1.280 | 1.540 | 3.300 |
| 14.000 | 1.340 | 1.630 | 3.420 |
| 14.250 | 1.370 | 1.740 | 3.580 |
| 14.500 | 1.400 | 1.820 | 3.750 |
| 14.750 | 1.450 | 1.920 | 3.920 |
| 15.000 | 1.480 | 2.020 | 4.080 |
| 15.250 | 1.510 | 2.100 | 4.220 |
| 15.500 | 1.530 | 2.200 | 4.360 |
| 15.750 | 1.540 | 2.280 | 4.480 |
| 16.000 | −9.990 | 2.360 | 4.610 |
| 16.250 | −9.990 | 2.450 | 4.730 |
| 16.500 | −9.990 | 2.530 | 4.820 |
| 16.750 | −9.990 | 2.630 | 4.960 |
| 17.000 | −9.990 | 2.720 | 5.120 |
| 17.250 | −9.990 | 2.810 | 5.250 |

**Note.** Isochrone derivation in Section 3.1.

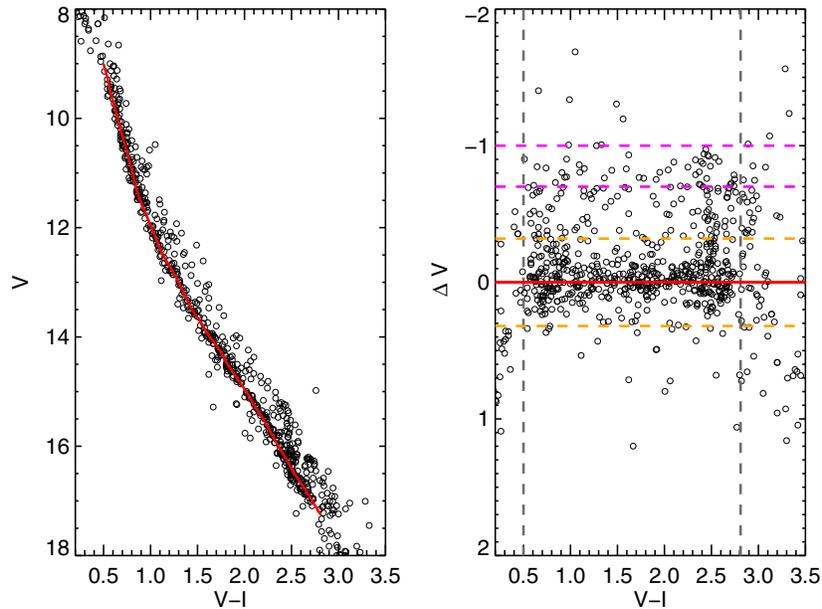

**Figure 3.** $V-I$ color–magnitude diagram (left) and color–displacement diagram (right) were used simultaneously to refine the 100 Myr single star locus by fitting a curve (red line) to the locus of stars that are scattered symmetrically about the line by ±0.3 mag. This technique is used to define an empirical isochrone for each of the color magnitude diagrams used in our analysis. Stars that remain consistently above the main sequence in all color–magnitude diagrams by $-0.4 > \Delta V > -1.0$ mag are flagged as a binary. The magenta line at 0.75 mag denotes the typical displacement for equal-mass binaries. Stars either consistently above the main sequence by 1.0 mag (upper magenta dashed line) or below by 0.3 mag (orange dashed line) are flagged as probable nonmembers.
(A color version of this figure is available in the online journal.)

### 3.2. Identifying Binaries and Nonmembers

To produce as clean a set of likely single stars as possible, we set a generous window of $-0.4 > \Delta V > -1.0$ to identify candidate binaries. Equal-mass binary systems are displaced by $\Delta V = -0.75$ mag in each CMD. Probable binaries are flagged if they lie within this range in two or three $V$ CMDs. We further verify binarity with $\Delta I$ values in the $I$ versus $I-K$ CMD, which is important for stars redward of $B-V = 1.45$ where $\Delta V$ values are not used. Nonmembers are defined as $\Delta V < -1.0$ mag or $\Delta V > 0.3$ mag consistently in all four CMDs and checked for additional evidence in the literature (such as objects with large proper motion errors).

To estimate binary detection sensitivity, synthetic binaries were generated using values from the S07 empirical isochrone. One set of 12 synthetic binary systems has a 1 $M_\odot$ primary with color and magnitude values of $V = 10.00$, $B-V = 0.560$, $V-I = 0.650$, $V-K = 1.378$, and $I-K = 0.728$. Another set of 12 synthetic binary systems has a 0.7 $M_\odot$ primary with color and magnitude values of $V = 13.00$, $B-V = 0.560$, $V-I = 1.31$, $V-K = 2.83$, and $I-K = 1.562$. Estimated masses for the primary star are from the mass versus $I-K$ relation of Baraffe et al. (1998). From these simulated binaries, the minimum detectable mass ratios to produce displacement values between $-0.4 > \Delta V > -1.0$ mag are $q = 0.62$ for a 1 $M_\odot$ primary and $q = 0.50$ for a 0.7 $M_\odot$ primary system. There are almost certainly binaries with more extreme mass ratios, but these are not detectable given our typical photometric errors and search criteria.

In this manner, 48 probable binaries and 14 likely nonmembers are identified in the new photometric set and are flagged in Table 1. Regardless of previously reported proper motion membership probabilities, the photometrically identified non-members were excluded from the rest of this study. Figure 4 highlights the identified nonsingle stars.

### 3.3. Displacement of Pleiades Single Stars Relative to ZAMS

In principle, it would be ideal to define a purely empirical ZAMS with which to compare the empirical Pleiades isochrone that we have defined above. Our first attempt at deriving an empirical ZAMS used the 650 Myr Praesepe cluster, an older analog to the Pleiades due to its richness (>1000 members; Kraus & Hillenbrand 2007) and proximity (181.5 pc; van Leeuwen 2007). However, the Praesepe photometric sample was not rich enough fainter than $V \sim 14$ in the $B$ and $V$ bands. A second attempt involved increasing the 650 Myr sample by adding photometry from the Hyades, since this cluster is better sampled at the lowest masses due its proximity (45 pc; Perryman et al. 1997). However, given that our program required a very accurate ZAMS, we felt that the scatter exhibited by the combined Praesepe and Hyades diagram was too large to allow us to proceed.

Potential reasons for the scatter could be distance uncertainties and/or metallicity effects. The Hyades is much closer than the Pleiades and Praesepe, and consequently one cannot assume the same distance modulus to all Hyades members. Accurate (*Hipparcos*) distances are known for the brighter Hyades members, but the faint members of most importance to our project either do not have *Hipparcos* estimates or have comparatively low accuracy parallax estimates. Also, both the Hyades and Praesepe have higher than average metallicity (and there is significant diversity in the published metallicity estimates for Praesepe). The effect of this on the CMD can be modeled (e.g., An et al. 2007), but it adds uncertainty to the resulting curve.

Therefore, instead of a completely empirical set of ZAMS relations as we had hoped to derive, we have decided to adopt a set of semi-empirical ZAMS relations created by VandenBerg & Clem (2003). Those authors combined the best available theoretical stellar evolutionary models with semi-empirical color–$T_{\rm eff}$ relations vetted by comparison to a wide range of observational data. Based on our own limited tests, we believe these are the best available ZAMS relations in terms of reproducing the real loci of low-mass, roughly solar metallicity stars for $BVI_cK$ photometric planes.

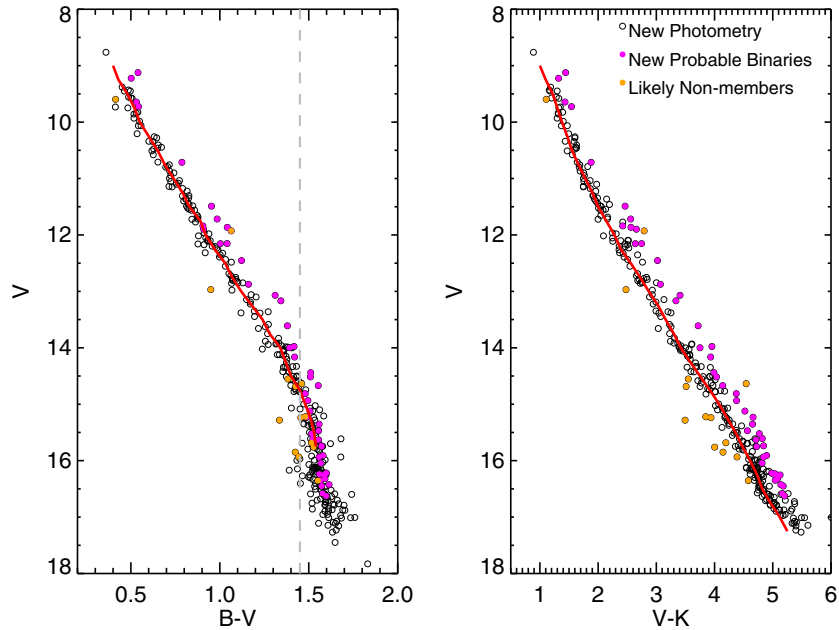

**Figure 4.** $B-V$ and $V-K$ color–magnitude diagrams of 48 probable binary systems (pink), 14 non-members (orange), and 288 single stars (empty circles) from the new photometry for 350 Pleiades stars. $B-V$ colors greater than 1.45 (dashed line) were not used in the search for reasons described in Section 3.1. Single stars are displaced between $\Delta V \pm 0.3$ relative to the newly defined empirical 100 Myr isochrone (red line) in all three $V$ magnitude CMDs and the $I$ vs. $I-K$ CMD (as described in Section 3.1). Binaries and nonmembers are flagged in Table 1.
(A color version of this figure is available in the online journal.)

In Figure 5, the identified Pleiades single stars are compared to the newly defined empirical 100 Myr isochrones (Section 3.1, blue curve) and to the semi-empirical ZAMS (VandenBerg & Clem 2003, green curve). The empirical relations clearly demonstrate that low-mass Pleiads are displaced relative to the "expected" ZAMS, as has been previously reported (e.g., Stauffer et al. 2003).

To quantify trends in the displacements with respect to broadband color, in Figure 6 each Pleiades star's $V$ magnitude displacement, $\Delta V$, is calculated relative to both the newly defined empirical 100 Myr isochrone (blue) and the semi-empirical ZAMS (green) in each CMD. $\Delta V = V_{\rm obs} - V_{\rm ms}$, where $V_{\rm obs}$ is the observed $V$ magnitude and $V_{\rm ms}$ is the expected $V$ magnitude from either the 100 Myr isochrone or the ZAMS. A negative $\Delta V$ corresponds to a displacement above the respective curves (i.e., brighter), and a positive value is below. Stars identified within $B-V$ color range of 1.0–1.45 are highlighted to show their displacement in each observational plane. This color range corresponds to mid K and early M stars (Kenyon & Hartmann 1995).

By construction, the stars show little to no displacements in the three CMDs relative to the newly defined empirical isochrones (blue symbols). In contrast, relative to the semi-empirical ZAMS (green circles), the mid K and early M dwarfs are positively displaced (i.e., they are below the ZAMS) in the $B-V$ CMD, and nearly coincident in the other two CMDs. The rest of the M dwarfs ($V-I > 1.9$ and $V-K > 4$) are displaced above the zero age main sequence (ZAMS) in the $V-I$ and $V-K$ observational planes.

### 3.4. Correlation between CMD Displacement and Stellar Rotation Period

We find that the low-mass Pleiades members have inconsistent displacements relative to the semi-empirical ZAMS in each observational plane. The nature of the $B-V$ displacements are consistent with those previously reported by Stauffer et al. (2003), who found that low-mass Pleiades members were displaced below (or blueward of) the ZAMS in the $B-V$ CMD. Also, Stauffer et al. (2003) found tentative evidence for a relation between the displacement in the $B-V$ CMD and stellar rotation for the small number of stars that had $v\sin i$ measurements. That small subset of stars were also found to be displaced above (or redward of) the ZAMS in the $V-K$ CMD.

In the time since Stauffer et al. (2003), photometric rotation period measurements have been reported for a large number of Pleiades members. Photometric rotation periods are preferable to $v\sin i$ measurements as the latter necessarily include unknown inclination effects. We successfully found matches between our sample and literature rotation periods for 277 Pleiades single stars from van Leeuwen et al. (1987), Stauffer & Hartmann (1987), Prosser et al. (1993a), Prosser et al. (1993b), Soderblom et al. (1993), Marilli et al. (1997), Krishnamurthi et al. (1998), Queloz et al. (1998), Terndrup et al. (2000), Messina (2001), Clarke et al. (2004), Scholz & Eislffel (2004), and Hartman et al. (2010). For a star with multiple observations, we averaged the reported rotation periods. Most multiply observed objects were observed twice (at most four times) and have rotation periods consistent at the 5%–10% level.

Stauffer et al. (2003) claimed that the displacement relative to the ZAMS in the $B-V$ CMD was spectral type dependent, with Pleiades G dwarfs having no significant displacement and stars later than K2 showing the effect. In order to directly compare to that work, we have estimated rough spectral types for all the

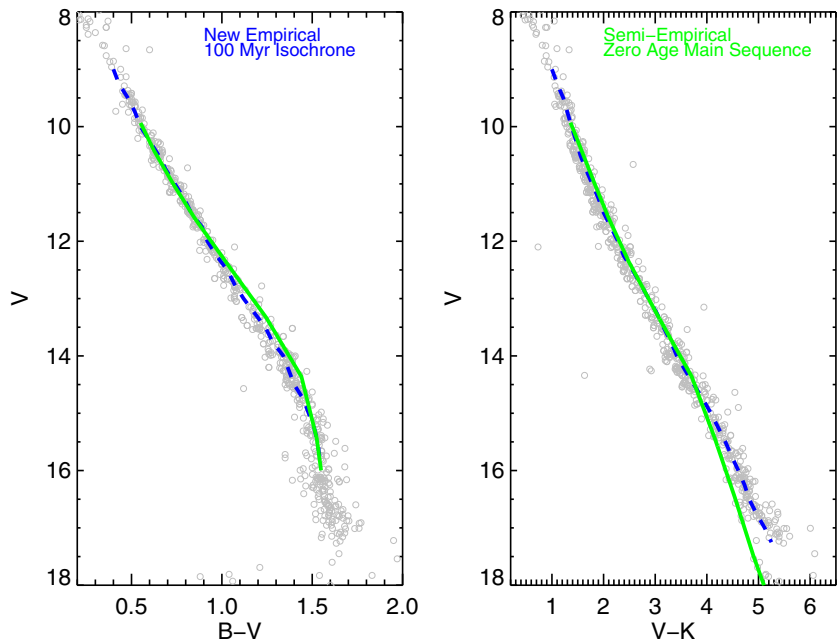

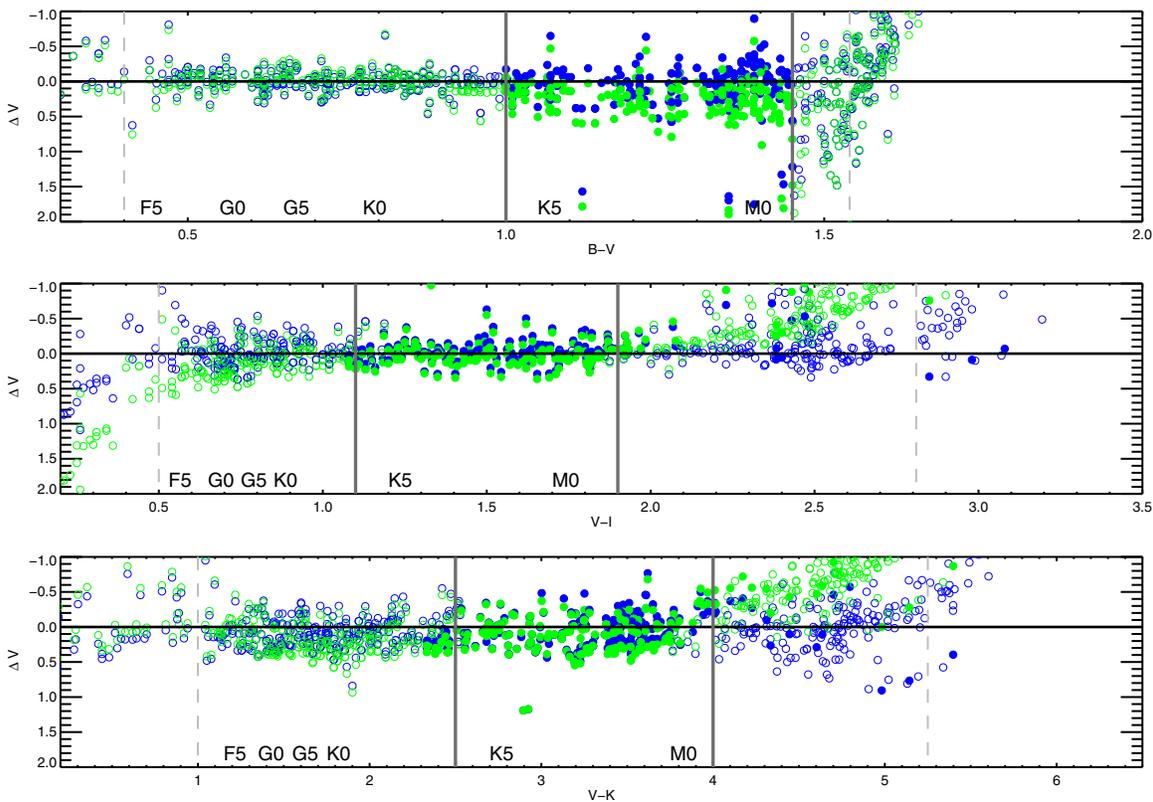

**Figure 5.** Color–magnitude diagrams of Pleiades single stars compared with the empirical 100 Myr isochrone newly defined in this study (blue curve) and the semi-empirical ZAMS (green curve). Pleiads with $12.5 < V < 15$ are displaced blueward in the $B-V$ CMD and single stars fainter than $V \sim 14.5$ are displaced redward of the ZAMS in the $V-K$ CMD, a trend that has been previously reported (e.g., Stauffer et al. 2003).

(A color version of this figure is available in the online journal.)

**Figure 6.** Displacement values vs. color of Pleiades single stars relative to the newly defined empirical 100 Myr isochrone (blue circles) and relative to the semi-empirical ZAMS (green circles). Negative displacement values correspond to stars above a given isochrone, positive displacement values correspond to stars below the isochrone. The dashed gray lines are the ends of the empirical 100 Myr isochrone. The solid vertical bars enclose the region of $B-V$ colors where the anomalous blue colors are present. The data points in this region are coded in filled circles. This same set of stars is highlighted in the other two plots, using again filled circles and transforming the color range for the vertical bars into equivalent $V-I$ or $V-K$ colors. The filled circles are late K and early M dwarfs within the cluster. In the $V-I$ and $V-K$ plots redward of $V-I = 2.3$ or $V-K = 4$, mid to late M Pleiades dwarfs are systematically displaced above the ZAMS; this is either in part or in whole due to the Pleiades stars still being on their PMS tracks.

(A color version of this figure is available in the online journal.)

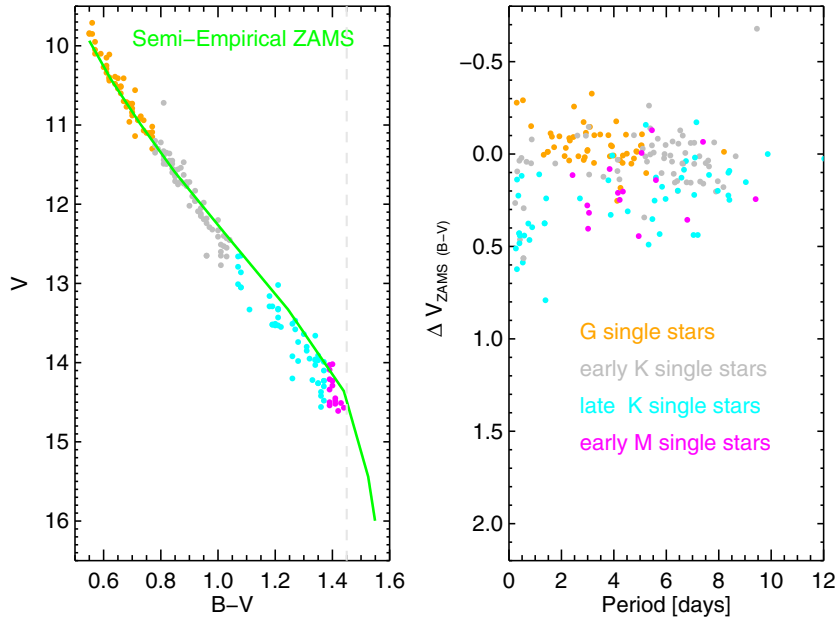

**Figure 7.** $B-V$ CMD (left) and rotation vs. displacement (right) of Pleiads with available literature rotation periods relative to the semi-empirical zero-age main sequence (blue). Spectral types G (orange), early K (K0 to K4; gray), late K (K5 to K7; blue), and early M (pink) single stars were categorized using the color–$T_{eff}$ relation from Kenyon & Hartmann (1995).

(A color version of this figure is available in the online journal.)

stars with rotation periods using their $V-K$ colors and based on the color-spectral type relation provided in Kenyon & Hartmann (1995). We subdivide the K dwarfs into early K (K0 to K4) and late K (K5 to K7). We exclude stars with $B-V$ color values greater than 1.45 for reasons mentioned in Section 3.1.

We compare the 277 Pleiades members with rotation periods in Figure 7. The left panel shows the stars coded by their spectral types in the $B-V$ CMD relative to the semi-empirical ZAMS (green curve). It is evident that the F and G type stars tend to follow the ZAMS closely, whereas the K and M type stars are systematically shifted blueward of the ZAMS. Figure 8 shows the distributions of $B-V$ CMD displacement values by spectral type. The right panel of Figure 7 relates these $B-V$ displacement values to the stars' measured rotation periods. It appears that the earlier type stars have a mostly scattered relation to rotation period, whereas the mid K and early M stars suggest a trend of increasingly positive (downward/blueward) displacement with shorter rotation period.

To quantify the likelihood of these possible trends, a non-parametric Spearman rho rank correlation test was applied (Figure 9, top). When applied to the full sample, the test returns a null-hypothesis probability of 29%, i.e., there is not a statistically significant correlation. However, subdividing the sample into F/G and K/M types, we find a null-hypothesis probability of 5.2% for the F/G stars and a probability of 0.008% for the K/M stars. In other words, the null hypothesis of no correlation is strongly rejected for the K/M stars with 99.992% confidence, in the sense that faster rotators (shorter periods) are displaced further below the ZAMS. A linear fit is applied for illustrative purposes only in Figure 9.

The same analysis was performed using the displacement values from the $V-K$ CMD (Figure 9, bottom). Here, the null hypothesis of no correlation can be rejected for both the F/G and K/M subsets with greater than 99.99% confidence. In this case, the sense is that the rapid rotators are more displaced upward/redward relative to the ZAMS.

## 4. DISCUSSION

We have verified, with high statistical significance, the suggestion of Stauffer et al. (2003) that for K and early M dwarfs ($1.0 < B-V < 1.45$) in the Pleiades rapid rotation causes stars to be displaced below the ZAMS in a $B-V$ based diagram, and above the ZAMS in a $V-K$ based diagram (Figure 9). This result is consistent with the interpretation that rapid rotation may induce strong enough magnetic fields to produce spots (manifesting as the redward excess in $V-K$) and plages (manifesting as the blueward excess in $B-V$). We furthermore are able to conclusively verify, with very high statistical significance, the suggestion of Stauffer et al. (2003) that these CMD displacements are connected to rotation effects (Figures 7 and 9).

The CMD displacements appear to be real physical effects intrinsic to the stars and not, for example, an artifact of foreground reddening, metallicity differences from star to star, or multiple episodes of star formation in the cluster. Although the southern part of the Pleiades sits behind a small CO cloud, less than a dozen of the members with accurate BVI photometry have significant excess reddening compared to the Av = 0.12 foreground reddening which is commonly accepted for the cluster. Because the reddening vector in the $V$ versus $B-V$ CMD is nearly parallel with the ZAMS, the excess reddening for these stars (Av's less than 1.0 mag except for one star) has no significant effect on our conclusions.

Within the Pleiades cluster itself, there is no evidence to date for a metallicity difference between the stars. A noncoeval star formation history could potentially lead to systematic displacements as was once proposed by Herbig (1962), but again this would not lead to opposite displacements in different CMDs. Most importantly, none of these explanations can account for the correlation between the magnitude of the CMD displacement with rotation.

Indeed, the correlation with rotation is an important clue to constraining the likely physical cause of the CMD displace-

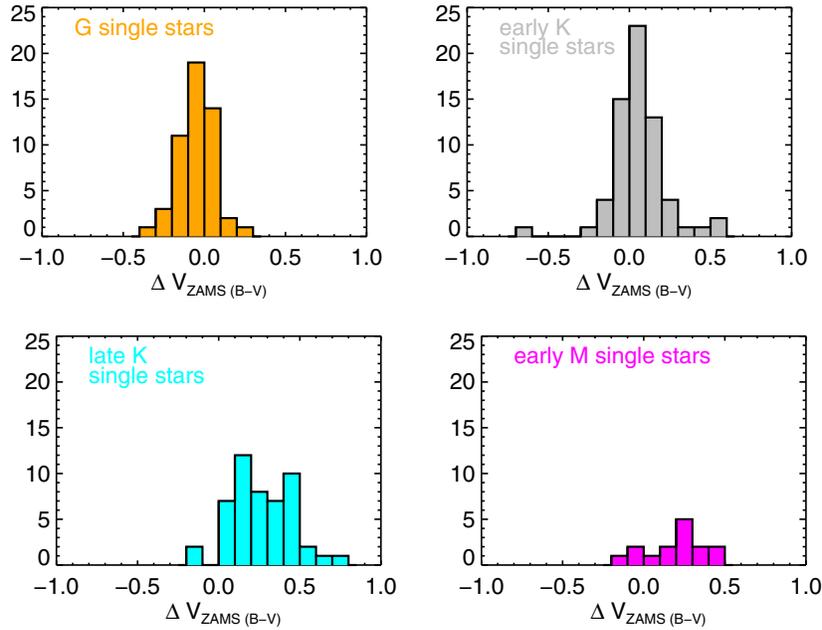

**Figure 8.** Distribution by spectral type of displacement relative to the semi-empirical ZAMS. Negative displacements are above and positive displacements are below the ZAMS. The trend is for G single stars to be centered on the ZAMS, early K dwarfs to have a slight displacement below, while the late K and early M single stars show a significant trend to be displaced below the ZAMS.

(A color version of this figure is available in the online journal.)

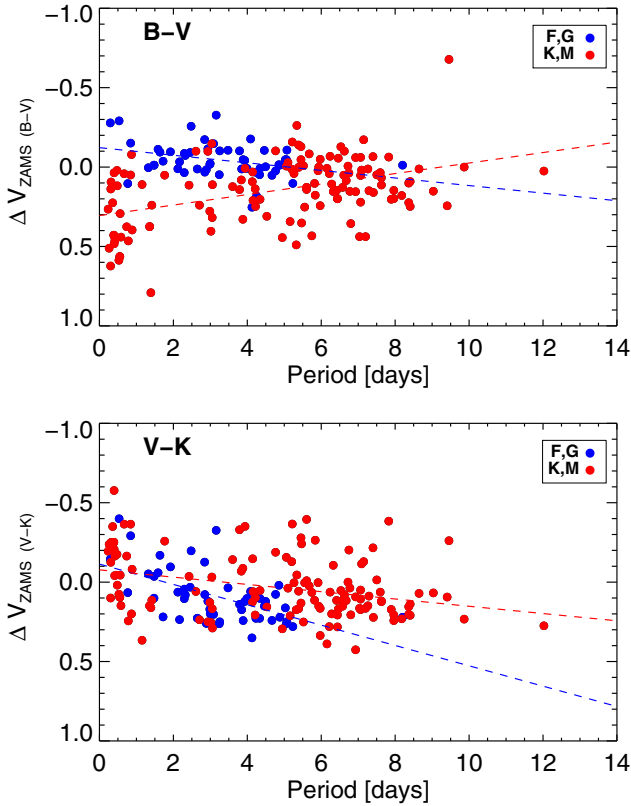

**Figure 9.** Top: correlation of displacement (relative to the semi-empirical ZAMS) vs. rotation period for stars in the $B-V$ CMD. F and G stars do not show a significantly significant correlation, whereas the K and M stars show a correlation at greater than 99.992% confidence in the sense that the rapid rotators are more displaced below/blueward of the ZAMS. Dashed lines are linear fits to guide the eye. Bottom: same as above but for stars in the $V-K$ CMD. Here, both F/G stars and K/M stars show a statistically significant at greater than 99.99% confidence in the sense that the rapid rotators are more displaced above/redward of the ZAMS.

(A color version of this figure is available in the online journal.)

ments. As suggested by Stauffer et al. (2003), an observational effect of a strong magnetic field is expected to be a temperature differential across the photosphere in the form of dark spots and bright plage regions. If magnetic activity manifesting as spots and plages is the cause, then this effect should in turn be connected with the stellar rotation, as stellar magnetic fields are in general thought to be induced by rapid rotation. Rapid rotation should simultaneously increase the strength of the field and the complexity of the field (degree of field multipolarity). Further modeling of the effects of magnetic fields on the inhibition of convection will be needed to better understand this mechanism physically.

## 5. SUMMARY AND CONCLUSIONS

In this study of the Pleiades, a benchmark open cluster at an age of 100 Myr, we have obtained new calibrated $BVI_C$ photometry for 350 proper motion members over the range of $8 < V \lesssim 17$, reported in Table 1. These new measurements roughly double the number of low-mass K and M type stars with high quality photometry in this cluster. In the new photometric catalog, 48 new probable binary systems and 14 likely noncluster members are identified. With the robust set of photometry at the low-mass end, the 100 Myr empirical isochrone has been refined for single star cluster members.

Using our larger photometric sample, we confirm the previous findings of Stauffer et al. (2003) of a systematic displacement of the low-mass stars to be bluer than expected when compared to the semi-empirically defined ZAMS of VandenBerg & Clem (2003). We tested the hypothesis from Stauffer et al. (2003) that the CMD displacements are driven by temperature inhomogeneities induced by magnetic fields from their rapid rotation. It was speculated that the physical process is rotationally driven dynamo creation of surface dark spots and hot plage regions, which can result in the observed blue excesses (plage) and red excesses (spots). We calculated stellar CMD displacement values for each star and have taken advantage of recently reported photometric rotation periods for large numbers of Pleiads, which was not available at the time of Stauffer et al. (2003).

In our work, there is a highly statistically significant correlation between rapid rotation and CMD displacement (Figure 9), essentially a correlation between rotation and the blue and red excesses in the spectral energy distributions of the stars, strongly corroborating these ideas.

B.L.K. is supported by the NSF Graduate Research Fellowship DGE-0909667. B.L.K. and K.G.S. acknowledge NSF PAARE AST-0849736. B.L.K. acknowledges the Science Support Consortium and MK&FC for their endorsement.